# TeV J2032+4130: a not-so-dark Accelerator?

Y. M. Butt[1], J. A. Combi[2], J. Drake[1], J. P. Finley[3], A. Konopelko[3],
M. Lister[3], J. Rodriguez[4], D. Shepherd[5]

**Abstract**

The HEGRA gamma-ray source TeV J2032+4130 is considered the prototypical "dark accelerator", since it was the first TeV source detected with no firm counterparts at lower frequencies. The Whipple collaboration observed this source in 2003-5 and the emission hotspot appears displaced about 9 arcminutes to the northeast of the HEGRA position, though given the large positional uncertainties the HEGRA and Whipple positions are consistent. Here we report on Westerbork Synthesis Radio Telescope (WSRT), Very Large Array (VLA), Chandra and INTEGRAL data covering the locations of the Whipple and HEGRA hotspots. We confirm a dual-lobed radio source (also see Marti et al., 2007) coincident with the Whipple hotspot, as well as a weak, partially non-thermal shell-like object, with a location and morphology very similar to the HEGRA source, in our WSRT and mosaicked VLA datasets, respectively. Due to its extended nature, it is likely that the latter structure is a more plausible counterpart of the reported very high energy (VHE) gamma-ray emissions in this region. If so, TeV J2032+4130 may not be a "dark accelerator" after all. Further observations with the new generation of imaging Cherenkov telescopes are needed to pin down the precise location and morphology of the TeV emission region and thus clear up the confusion over its possible lower frequency counterparts.

[1] *Harvard-Smithsonian Center for Astrophysics, 60 Garden St., Cambridge, MA 02138, USA*

[2] *Departamento de Física (EPS), Universidad de Jaén, Campus Las Lagunillas s/n, 23071 Jaén, SPAIN*

[3] *Department of Physics, Purdue University, West Lafayette, IN 47907, USA*

[4] *CEA Saclay, DSM/DAPNIA/SAp,F-91191 Gif sur Yvette, FRANCE*

[5] *NRAO, P.O. Box O, Socorro, NM 87801-0387, USA*

## Introduction

In 2002 the HEGRA collaboration reported on a steady and extended (~6.2 arcmin radius) TeV gamma-ray emitting region, TeV J2032+4130, near the massive stellar association Cygnus OB2 (Aharonian et al., 2002, 2005). We followed-up this source using the Very Large Array (VLA) and CHANDRA, but found no obvious counterparts at the lower frequencies (Butt et al., 2003, 2006; Mukherjee at el., 2003, 2006). Paredes et al. (2007) and Marti et al. (2007) have recently reported on some deeper radio observations which do show a handful of very interesting potential counterparts. Several possible origins of the gamma-ray emission have been suggested in the literature having to do, variously, with the termination lobes of Cyg X-3 (Aharonian et al., 2002; see also, Marti, Paredes & Peracaula 2000); the stellar winds in Cyg OB2 (Aharonian et al., 2002; Butt et al., 2003; 2006; Domingo-Santamaria & Torres, 2006); a possible proton-blazar (Mukherjee et al., 2003); a pulsar wind nebula (Bednarek 2003; 2006); or possible microquasars (Paredes et al., 2007). However, since none of these possible scenarios has yet been conclusively verified, this, and other similar TeV sources without firm counterparts, have been dubbed "dark accelerators" by some authors. (For a recent overview see, e.g., Aharonian et al, 2007). It is commonly believed that they are most likely of hadronic origin due to the weak emission in other wavebands. TeV J2032+4130, being the first discovered is, in fact, the prototype of this informal class of sources.

Using archival 10m data from 1989-90, the Whipple collaboration confirmed the existence of TeV J2032+4130 (Lang et al., 2004), and more recently a total of 65.5 hr of Whipple data taken during 2003-2005 towards this source has been reported (Konopelko et al. 2007). The analysis of the latest dataset reveals a distinct excess in the field of view at a significance of 6σ. The centroid position of this γ-ray source is $\alpha=20^h32^m27^s$, $\delta=41°39'17''$ and the estimated integrated flux is about 8% of the Crab-Nebula flux. The data are consistent with a point-like source (at the Whipple resolution), even though an extended source with an effective radius less than 7′ cannot be ruled out. The center of gravity of the γ-ray emission as seen in the latest Whipple dataset is about 9 arcmin northeast of that reported by HEGRA. However, given the statistical and systematic uncertainties in the source localization, which are ~4′ and 6′, respectively, the Whipple hotspot location can be considered consistent with the HEGRA γ-ray source position [Fig. 1a].

## Radio Analysis

*Westerbork Synthesis Radio Telescope (WSRT) data*

We reanalyzed the WSRT data of Setia-Gunawan et al. (2003; hereafter SG2003) using standard procedures with the AIPS package of the National Radio Astronomy Observatory (NRAO). SG2003 carried out a large mosaic of the Cygnus region comprised of a total observing time of six 12 hrs sessions spanning two observing periods. Two 12 hrs sessions were carried out in March–April 1996 and the remaining four 12 hrs were done in August–October of 1997 (SG2003). For these observations the WSRT beam size at 1.4GHz and 350 MHz was 13"×19" and 48"×75" respectively. As reported by SG2003, the limiting 5-$\sigma$ flux-densities were ~2mJy at 1.4GHz and ~10mJy at 350MHz. We convolved the radio map at 1420 MHz with a similar synthesized beam as the 350 MHz observations in order to perform the spectral index measurements.

We have previously noted two non-thermal radio sources with a jet-like structure marginally coincident with the HEGRA position in the WSRT data reported by SG2003 at 350 MHz and 1.4 GHz (Fig 1; see also Table 3 in Butt et al., 2006; hereafter B2006; and, Marti et al., 2007). We rechecked the intensities and spectral indices of the dual-lobed radio structure in WSRT data and confirm that they agree with the values stated by SG2003 (Table 1).

| # (SG2003) | Position (J2000) | 1.4GHz Flux Density | 350MHz Flux Density | Sp. Index |
|---|---|---|---|---|
| 217 | 20 32 1.22, +41 37 13.64 | 36±5 mJy | 87±5 mJy | -0.65 |
| 218 | 20 32 2.16, +41 37 59.24 | 39±5 mJy | 122±3 mJy | -0.84 |

**Table 1: Characteristics of the non-thermal dual-lobe radio object found in WSRT data (SG2003). The spectral index is defined as $S_\nu \propto \nu^\alpha$.**

There is also a weak diffuse non-thermal radio condensation roughly aligned with the axis of the lobes located ~5 arcmin further SW of the dual-lobed jet-like source, but it is unknown whether the structures are related (Fig. 2). This radio source is located at $\alpha_{2000}, \delta_{2000}$ ~ (20 31 53, +41 32 35) and has a 350MHz flux of ~11.6mJy (i.e. above ~5-$\sigma$ limiting flux threshold). At 1400MHz, the source is so weak that it is only marginally above the 5-$\sigma$ limiting flux density level of ~2mJy. The spectral index is of this radio emission is –1.3±0.5. At least two other non-thermal radio sources are also found in WSRT data in the extended radio field, one of which is also dual-lobed. However, the locations of these sources are inconsistent with the position of either one or both TeV gamma-ray hotspots, and therefore we do not consider them likely counterparts of the TeV emission (Fig. 1). Marti et al. (2007) have also examined these compact sources in further detail using Giant Metrewave Radio Telescope (GMRT) and VLA C-configuration observations and also find that they are unlikely to be the counterparts of the TeV source.

To estimate the total diffuse radio flux in the extended radio emission region coincident with the Whipple hotspot, we considered all the radio flux with the elliptical region shown in Fig. 1, except for that due to the two enclosed compact radio sources. We found the following flux densities at the two frequencies for the diffuse radio emission in this region: F(350 MHz)= 2.1±0.3 Jy and F(1420 MHz)= 2.8±0.4 Jy which is indicative of a thermal source.

*VLA data*

On April 29 2003 we had also carried out a mosaic observation of the TeV J2032+4130 source region with the VLA[6] at both 6 and 20cm. The array was in the D-configuration and the flux calibrator was 3C48 and the gain calibrator, J2007+404. The source fields were observed with 3.5' to 6' spacing in a 5-point pattern, the primary beam of the VLA at 6cm being about 9'.

The resulting 6cm data were reduced with the Common Astronomy Software Applications (CASA) software and imaged with Astronomical Information Processing Software (AIPS++). The final mosaic image was weighted with robust weighting that provided an optimal compromise between point source sensitivity and minimum noise levels. (A Briggs robustness factor of 0.5 was used.) The image was gridded to form a single mosaic and then deconvolved with a CLEAN algorithm that allowed multiple scales to be used for the clean components to maximize sensitivity to extended structure while still preserving the image flux density[7]. The resulting image (Fig. 3) has uniform sensitivity across the mosaic field, however the flux density decreases toward the edge of the field as the primary beam response decreased.

---

[6] NRAO VLA proposal AB1075 (PI: Butt) Observed April 29, 2003. 10 hours total in a five-pointing mosaic observation at 6 and 20cm. The NRAO is a facility of the NSF operated under cooperative agreement by Associated Universities, Inc.

[7] Specifically, scale sizes of 0", 12" and 40" were used. Note that Marti et al. (2007) consider the central 6cm VLA pointing of this dataset in their image analysis (their Fig. 3). The present analysis presents the full co-added mosaic image of the 5 VLA D-configuration pointings at 6cm (our Fig. 3).

The flux was measured in an image that was corrected for the mosaic primary beam response so the flux density was constant while the noise increased toward the edge of the mosaic field. The extended structure has a flux density of approximately 414±12 mJy at 6cm. In comparing the 6cm data with the 20cm as analyzed by Paredes et al. (2007) [also in D-configuration], it is evident that the western region of this extended shell-like structure is predominantly non-thermal, as also borne out in the analysis of Marti et al. (2007). This structure is centered approximately at $\alpha_{2000}, \delta_{2000}$= (20 31 55, +41 29 00) with a radius of about 5 arcmin, consistent with the reported location and extension of the HEGRA source. The emission is weak and at 1.4GHz only the western part of the full ring-like structure is evident (Fig. 3). The radio spectral index is variable across the structure and ranges from $\alpha \sim$ –0.7 in the west, to ~ +0.6 or larger in the east. We can thus tentatively say that the western region of this object is apparently more non-thermal than the rest of the structure. This radio shell structure is not evident in the 350 MHz WSRT data due to its weak intensity – it is below the noise level.

## Chandra

A Chandra X-ray Observatory ACIS-I observation of the region in the vicinity of TeV J2032+4131 was described by B2006. Here, we reanalyzed the same 48728s exposure. Analysis was performed using CIAO software Version 3.3.0.1. B2006 listed sources found by an automated wavelet-base source detection algorithm "*wavdetect*", and three sources from that list that fall close to the radio source: 36 (**E** in Fig. 1), 153 (**F** in Fig.1) and 227 (**G** in Fig. 1; this is the same as source 1 of Mukherjee et al., 2003). Of these, 36 and 153 are faint while 227 is a bright source with a count rate of about 0.015 count/s. In addition to these sources, we also located by eye two additional regions of enhanced X-ray brightness that appear to be slightly more extended than a point source due to the effects of point spread function (psf)-distortion this far off-axis. One of these falls between the two radio lobes but is off-set slightly east of their centre; the other is coincident with the southern radio lobe. These two regions of enhanced X-ray brightness can each be described by two positions derived from a 8× binned image of the field: Central (**A** & **B** in Fig. 1) , with $\alpha_{2000}, \delta_{2000}$=(20 32 02.4, +41 37 37) and $\alpha_{2000}, \delta_{2000}$=(20 32 02.1, +41 37 34); and the southern-lobe (**C** & **D** in Fig. 1), with $\alpha_{2000}, \delta_{2000}$=(20 32 01, +41 37 10.6) and $\alpha_{2000}, \delta_{2000}$=(20 32 00.6, +41 37 06.7). We find that both these regions contain more counts in the 2-10keV band than in the 0.5-2keV one. For a power-law photon spectrum, this division of photons suggests an index of $\Gamma \lesssim 1$, implying a rather hard nature for these sources.

The diffuse non-thermal radio condensation located ~5 arcmin SW of the dual-lobed source also has two X-ray sources coincident with its location: SW Condensation X1: $\alpha_{2000},\delta_{2000}$=(20 31 52.71, +41 32 40.20) and SW Condensation X2: $\alpha_{2000},\delta_{2000}$=(20 31 52.49, +41 32 09.43). The first is likely the X-ray counterpart of star #157 of Massey & Thompson (1991) at $\alpha_{2000},\delta_{2000}$= (20 31 52.63, +41 32 40.7), whereas the second could truly be a genuine X-ray counterpart of the radio condensation. The X-ray event characteristics of all these sources nearby or coincident with the relevant radio structures are listed in Table 2.

**Table 2: Characteristics of Chandra sources detected nearby or coincident with the non-thermal dual-lobe radio object. Note that the lobe sources are both rather hard in X-rays.**
*\* total counts in source region (ie source + background)*

| X-ray Source | Cts* | S/N | Frac. 0.5-2.5 keV | Frac. 2.5-10 keV | X-ray Flux (erg cm$^2$ s$^{-1}$) |
|---|---|---|---|---|---|
| Central: **A&B** | 32 | 2.8 | 0.28 | 0.72 | ~1.8× 10$^{-14}$ |
| South-lobe: **C&D** | 47 | 2.6 | 0.38 | 0.62 | ~2.6× 10$^{-14}$ |
| SW condensation X1 | 9 | 2.9 | 0.89 | 0.11 | ~5× 10$^{-15}$ |
| SW condensation X2 | 9 | 2.8 | 0.56 | 0.44 | ~5× 10$^{-15}$ |
| B2006:36   **E** | 24 | 2.3 | 0.38 | 0.62 | ~1.3× 10$^{-14}$ |
| B2006:153  **F** | 29 | 2.5 | 0.48 | 0.52 | ~1.6× 10$^{-14}$ |
| B2006:227  **G** | 718 | 26.0 | 0.31 | 0.69 | ~4× 10$^{-13}$ |
| Background | - | - | 0.29 | 0.71 | - |

## **Infrared, Optical and Hard X-ray Bands**

Near the center of the dual-lobed radio structure, and almost coincident with the X-ray emission of sources **A** & **B**, lies a 2MASS infrared source, 2MASS 20320186+4137377, with standard magnitudes J=16.401±0.107, H=15.325±0.096 and K=14.879. With the present available information, however, it is impossible to tell if this object is stellar or not, since neither a spectrum nor accurate reddening information is available. Marti et al. (2007), in fact, find that the near-IR source is offset slightly NW from the center of the dual-lobed structure and thus may be unrelated. There are also several other 2MASS sources nearby or coincident with the radio lobes, as shown in Fig 1b, but given the high stellar density it is also unclear whether these are related to this source.

We inspected the USNO-B1.0 catalog (Monet et al., 2003), but found no objects coincident with the central 2MASS source at optical wavelengths. However there is a star at: $\alpha_{2000},\delta_{2000}$= (20 32

00.1, +41 37 14) with $B_{magn}$=15.46, $V_{magn}$=14.00 (Star #179 in Massey & Thompson 1991), marginally coincident with the southern lobe of the radio structure. This star is also designated GSC 03161-00887.

We also reduced INTEGRAL/IBIS hard X-ray data with the Off line Scientific Analysis (OSA) software package version 6.0. We focused on the data from the IBIS Soft Gamma-ray Imager (ISGRI, Lebrun et al. 2003) which has the highest sensitivity and best angular resolution up to ~200 keV. We extracted images in four energy ranges, namely 20-40 keV, 40-80 keV, 80-150 keV and 150-300 keV, from all public science windows (scw, i.e. INTEGRAL pointings) that were less than 15 degrees from the Galactic Black hole Cyg X-1. This resulted in 1565 good scw between revolutions 12 and 218, for a total exposure time of about 3 Ms. None of the very high energy sources was spontaneously detected by the software. The presence of bright sources such as Cyg X-1 and Cyg X-3 in the field can be a cause of artifacts (fake sources) in mosaics of such a long accumulation time. We thus verified that the TeV J2032+4130 location did not fall on any such artifact. We estimated $3\sigma$ upper limits from the values of the variance at the gamma-ray position and the efficiency of the source detection in each energy range, following Upper Limit = 3 var$^{1/2}$/Efficiency. We found no positive detection of TeV J2032+4130 in the 20-300keV band and derived the following $3\sigma$ upper limits: 20-40 keV : 0.40 mCrab; 40-80 keV : 0.85 mCrab; 80-150 keV: 2.2 mCrab; 150-300 keV: 50 mCrab.

## **Discussion & Conclusions**

Motivated by the more northerly location of the center-of-gravity of Whipple hotspots (Lang et al. 2004; Konopelko et al., 2007) for the source TeV J2032+4130, as compared with the HEGRA one (Aharonian et al., 2002; 2005) we undertook a multiwavelength study of that region in pursuit of possible lower frequency counterparts of the reported high-energy emission. The main intent of the present report was to synthesize the findings in gamma-ray, hard & soft X-ray, optical, IR and radio bands with an eye to drawing some conclusions regarding the possible counterparts.

Detailed studies of this region in the radio and IR band have also been presented by Paredes et al. (2007) and Marti et al. (2007). Although the same 6 cm D-configuration VLA dataset has also been analysed by Marti et al. (2007), the main difference is that here we have carried out a full mosaicking analysis of the 5 VLA pointings in order to construct an accurate and wide field-of-view 6cm image (Fig. 3). Below we describe the data on the 3 possible counterparts: the non-thermal dual-lobed radio source; the radio condensation ~5 arcmin SW of that source; and, lastly, the diffuse radio emission detected in the VLA 6cm D-configuration observations.

*a) Dual-lobed Source*

We find a dual-lobed non-thermal radio object consistent with the position of the Whipple hotspot, and to a lesser extent, also the HEGRA TeV emission region. There is a 2MASS infrared source (2MASS J20320186+4137377) and coincident X-ray emission (sources **A** & **B**) near the center of this radio structure. The latter could be a signature of X-ray emission from nearby the putative compact object or disk emission. At least the southern radio lobe also appears to emit in X-rays, and quite possibly the northern one also. Both the central and southern lobe X-ray sources appear to have a hard spectrum, such as may be expected from an absorbed source, but the statistics are too poor to make any definitive judgment (Table 2).

With the information available we cannot yet make a definitive statement on the nature of the dual-lobed jet-like non-thermal radio source. If Galactic, it could well be a microquasar (eg. Bosch-Ramon, Aharonian and Paredes, 2005; Paredes 2006) or, less likely, pulsar-related (eg. Bednarek, 2006); if extragalactic, it is most likely a radio galaxy with powerful dual radio cocoons. M87 and Centuarus A are examples of TeV- and GeV-emitting non-blazar radio galaxies, respectively (eg. Grindlay et al., 1975; Sreekumar et al., 1999; Beilicke et al, 2005; Combi et al., 2003); however, at least in the case of M87, the VHE emission appears to be point-like and rapidly variable (Aharonian et al., 2006). On longer, year-long, timescales, however, M87 appears to be more steady. To better discriminate between the Galactic *vs.* extragalactic alternative one would need to detect the central object at optical and/or IR frequencies and obtain a spectrum to be able to test for the possible redshift of the H-alpha or other emission lines. If no redshift is found it would of course imply that this source is Galactic, possibly similar to a number of highly-absorbed INTEGRAL sources recently discovered (eg. Filliatre & Chaty, 2004; Chaty & Rahoui, 2006). In fact, Marti et al. (2007) have carried out such observations using the OMEGA2000 camera on the Calar Alto 3.5m telescope but have found no near-IR counterpart. We concur with them that this object is most likely a double-double radio galaxy. Interestingly, even if this radio source is unrelated to the gamma-ray emission reported by Whipple and HEGRA, it would still make an excellent candidate counterpart of the flaring TeV emission reported by the Crimean Astrophysical Observatory approximately 0.7° North of Cygnus X-3 (Neshpor et al., 1995).

*b) Weak diffuse source 5arcmin SW*

There is also a weak diffuse non-thermal radio condensation roughly aligned with the axis of the lobes and located about 5 arcmin SW [$\alpha_{2000},\delta_{2000}$= (20 31 53,+41 32 35) with a 350MHz flux of ~11.6mJy, see Fig. 2], but it is unknown whether the structures are related. This source is notable since it is located within the HEGRA emission region, and is also diffuse. If this source is a TeV emitter, possibly in addition to the suggested dual-lobed jet-like object, then it is possible that the extension reported by HEGRA (and not ruled out by Whipple) arises from the composite nature of these two emission regions. There is some X-ray emission coincident with the radio condensation, but this may well be a purely chance occurrence given the large number of X-ray sources in the field.

*c) Extended emission detected in the 6cm VLA D-configuration dataset*

Our mosaicked VLA data at 4.8 GHz also hints at the presence of weak (414±12 mJy at 4.8GHz) possibly shell-like emission with a radius of about 5arcmin which is consistent with the 1$\sigma$ extension reported by HEGRA for TeV J2032+4130 (Fig. 3). However, it is uncertain if this is one coherent object and could simply be due to selection effect of the extent of the limited mosaicked field: i.e. similar very weak diffuse radio emission may be present all over Cygnus OB2 and this structure appears to have been picked up due to the limited field of view. Additional VLA observations of a wider field are being planned and would certainly help clear up the possible association of the reported diffuse radio emission with the TeV source. The center of the radio emission in Fig. 3 (assuming it is a single coherent object for the moment) is approximately $\alpha_{2000},\delta_{2000}$= (20 31 55, +41 29 00), consistent with the HEGRA report and at least its western part appears to be non-thermal. Using the 20 cm VLA dataset Paredes et al (2007) have used equipartition arguments to infer that the diffuse radio structure may have an energy content of ~$10^{46}$ ergs and that it may be an efficient accelerator of non-thermal particles. If the radio emission is the signature of an SNR co-located with Cyg OB2, at a distance of 1.7 kpc, its radius would be only ~3pc and its age as low as ~500 years. Note that the young SNR G347.3-0.5 (RX J1713-3946) also is TeV bright and radio-dim (Lazendic et al., 2004; see also, Tang & Wang 2005). Alternatively, this radio structure may also be due to multiple large-scale shocks created by the cumulative action of the very powerful stars in Cyg OB2 (eg., Cesarsky & Montmerle, 1983). In this case, the radio emission would likely extend beyond the limits of the moasicked field shown in Fig. 3. It remains to be seen whether the intriguing, dominantly non-thermal, radio emission shown in Fig. 3 is physically related to TeV J2032+4130. Deeper, wider-field and more sensitive observations of this region at a handful of radio frequencies would be very useful to extract higher-fidelity data on this interesting structure(s), and are being planned.

Besides the dual-lobed source, relatively strong extended radio emission is also present within the area of the Whipple hotspot. Though this could also conceivably be a radio counterpart of the TeV emission, this extended radio emission is thermal in nature, with a mean index of +0.12±0.1. Furthermore, the extended thermal radio region lies well outside the HEGRA TeV J2032+4130 location. For these two reasons we disfavor its direct association with TeV J2032+4130.

We have discussed above the possible association of the dual-lobed radio structure, together with its plausibly related radio condensation 5 arcmin to the SW and the possible shell-like emission region, with TeV J2032+4130. However, we emphasize that the case is far from closed. The alternate associations of this VHE source with possible microquasars (Paredes et al., 2007) or with an outlying sub-group of very massive and powerful stars in Cyg OB2 are also somewhat persuasive (see, eg., Fig 1 in Butt et al., 2003 and Fig. 3 in B2006). In fact, Anchordoqui et al. (2006a, 2006b) have recently proposed an interesting new hypothesis for the origin of the high energy flux in the latter scenario: they suggest that the TeV gamma-rays could result from the photo-deexcitation of PeV energy nuclei that are themselves the photo-disintegration products of heavier nuclei broken-up in a bath of intense ultraviolet photons, such as would be present in Cyg OB2. Other interesting phenomena, such as possible time-correlated supernovae of a previous generation of stars in the Cyg OB2 region – which may have led to the ejection of sets of binary pulsars (Vlemmings, Cordes and Chatterjee 2004) – may also be involved in untangling the true scenario of the origin of the TeV flux seen here. Furthermore, a proper understanding of TeV J2032+4130 may also need to take into account the recent results from the air-shower arrays MILAGRO and TIBET, which have shown the entire Cygnus region to be a bright source of TeV gamma-rays (Atkins et al., 2005) and possibly also cosmic rays (Amenomori et al., 2006; Abdo et al.,2007).

Naturally, we are predisposed to believe that the counterpart of the extended TeV emission would also appear extended in nature and agree with Marti et al (2007) who also disfavor the association the TeV emission with any one of the individual compact radio sources[8].

---

[8] However, note that it is possible that several weak point-like TeV sources in this region are conspiring to appear as a single extended TeV source due to the limited psf of the Cherenkov observatories. Though it is unlikely that extragalactic sources are thus clustered, it is conceivable that the many stars in the outlying OB sub-group of Cygnus OB2 are in fact individually the TeV sub-sources and meld together to form what

Clearly, it would be very helpful to better determine the true location and morphology of the TeV gamma-ray emission region (TeV J2032+4130) with higher precision, and especially to confirm its extended nature. This is something that the new generation of Imaging Cherenkov telescopes are ideally suited for doing. We estimate, for instance, that just 20hrs of VERITAS time could yield a 6 sigma detection which would allow one to properly localize this source, and, simultaneously, give better morphological information possibly confirming or denying its association with the various possible lower frequency counterparts in the region.

---

appears as a extended TeV source to the Cherenkov observatories (e.g. Butt et al., 2006; Domingo-Santamaria & Torres, 2006).

## Acknowledgements

We thank an anonymous referee for useful comments which significantly improved the manuscript. YMB is supported by NASA/Chandra and NASA/INTEGRAL GO Grants and a NASA LTSA Grant. JPF and AK thank the US DOE for their continued support. JAC is a researcher of the program Ramon y Cajal funded jointly by the Spanish Ministerio de Ciencia y Tecnologia and Universidad de Jaen. J. C. also acknowledges support by DGI of the Spanish

Ministerio de Educacion y Ciencia under grants AYA2004-07171-C02-02 and FEDER funds and Plan Andaluz de Investigacion of Junta de Andalucia as research group FQM322. The National Radio Astronomy Observatory is a facility of the National Science Foundation operated under cooperative agreement by Associated Universities, Inc.

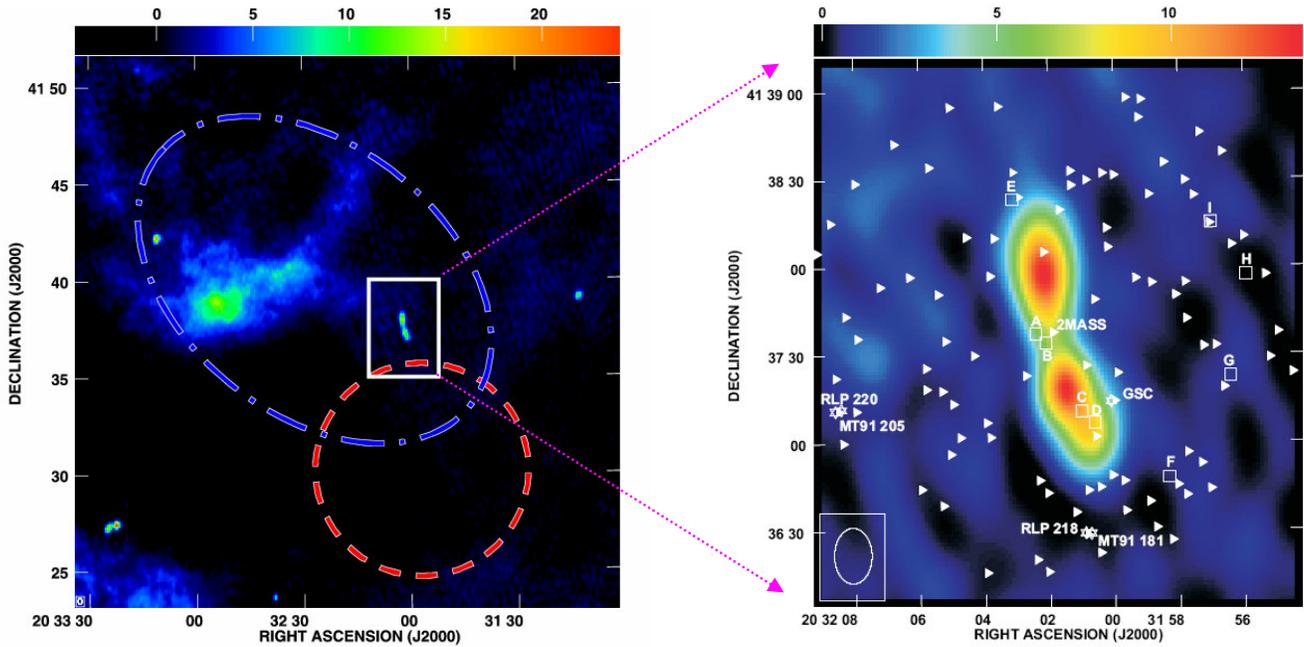

Figure 1: (**left**) 1400MHz WSRT map showing the approximate location of the Whipple TeV $\gamma$-ray hotspot (blue) as well as the extended HEGRA emission region (red). Note that the Whipple hotspot does not necessarily correspond to an extended TeV gamma-ray emission region but is simply the highest likelihood region of $\gamma$-ray emission, of either a point-like nature or an intrinsically extended region of radius less than ~7 arcmin. The blue ellipse shown corresponds roughly to the 600 excess counts level in Fig 3 of Konopelko et al. (2007). The non-thermal dual-lobed radio source discussed in the text is enclosed within a white rectangle. The other two non-thermal radio sources, WSRTGP 2031+4116 (also dual-lobed; $\alpha_{2000},\delta_{2000}$=20 33 23.19 +41 27 23.7), and WSRTGP 2031+4131 ($\alpha_{2000},\delta_{2000}$=20 33 10.6 +41 42 11) can be seen towards the lower-left and upper left regions of the figure, respectively. The large region of diffuse radio emission towards the center of the Whipple hotspot location is thermal in nature (see text). (**right**) A close-up of the dual-lobed non-thermal radio source of interest at 1.4GHz, overlaid w/ CHANDRA (□) and 2MASS (▸) counterparts. CHANDRA sources have been labeled alphabetically. Stars are shown as (✡) and appear with their catalog names. GSC refers to the star GSC 03161-00887.

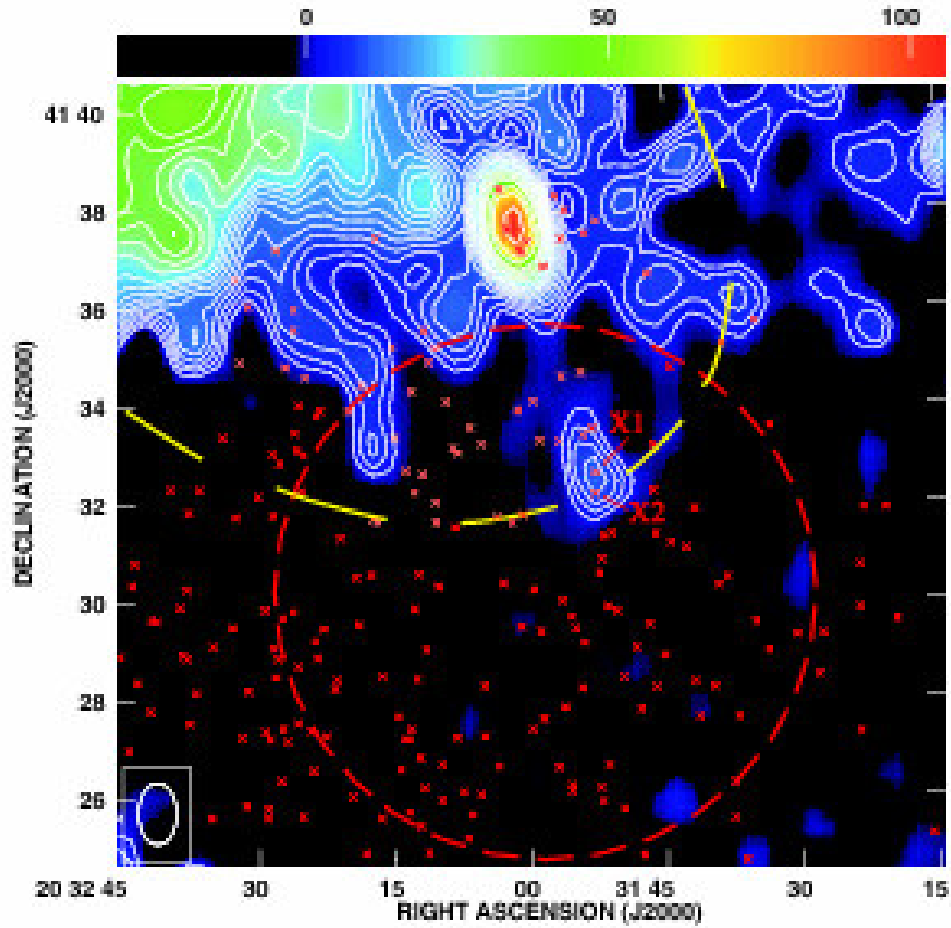

Figure 2: 350MHz WSRT data showing the diffuse non-thermal radio condensation at $\alpha_{2000}, \delta_{2000}$= (20 31 53,+41 32 35) which is aligned with the radio-lobes and has a flux of ~11.6mJy at 350 MHz. Since the resolution at 350 MHz is poorer than at 1420MHz the two lobes are not discernable in this image. CHANDRA X-ray sources are shown as red crosses, and the HEGRA (red) and Whipple (yellow) TeV emission regions are marked. Note that the CHANDRA field of view only covered part of the radio field displayed above.

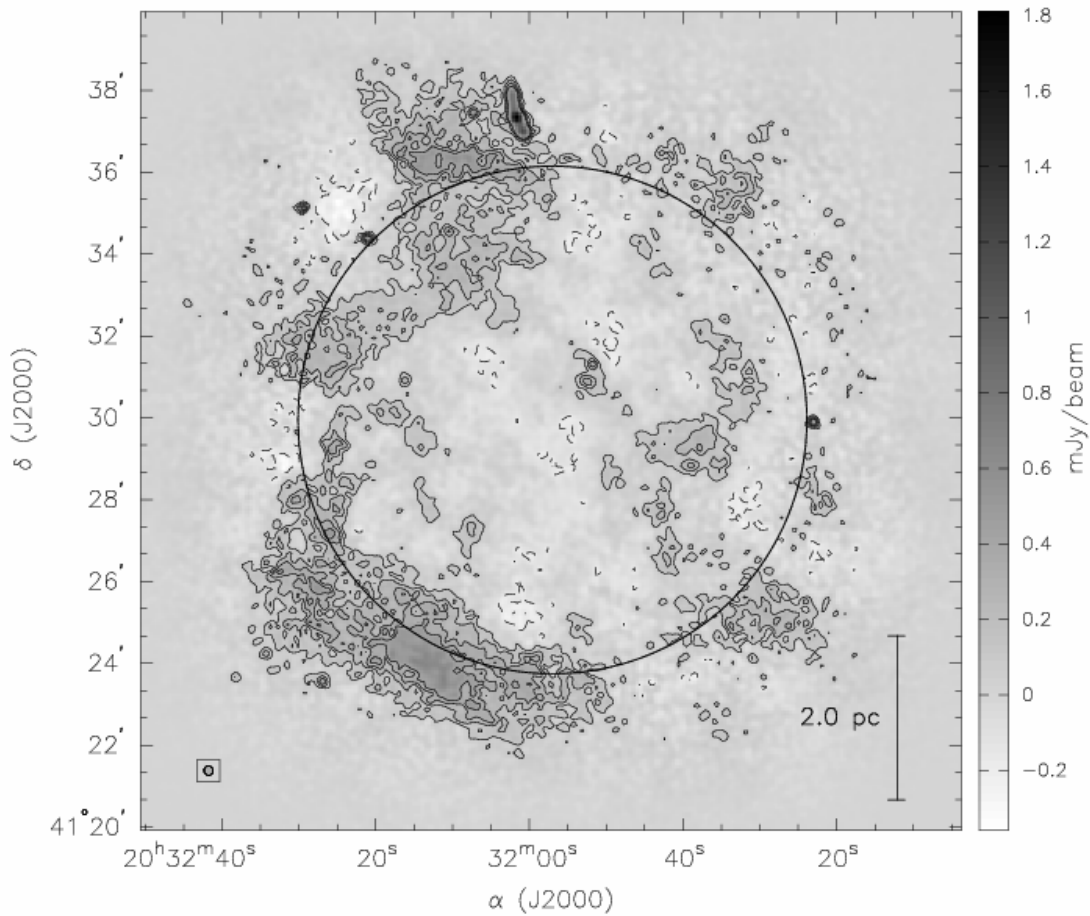

Figure 3: A mosaicked VLA image using 5 pointings at 4.8GHz (6cm) in D-configuration taken 29 April 2003. The presence of a possible weak roughly ring-like structure(s) with center approximately $\alpha_{2000}, \delta_{2000}$= (20 31 55, +41 29 00) and radius ~5arcmin, consistent with the dimensions of the TeV source reported by HEGRA (indicated by the overlaid circle), is evident (see text). The dual-lobed radio source can also be seen to the north as well as the close-by compact sources VLA-N and VLA-S of Paredes et al. (2007) near the center of the circle. The spectral index of the structure varies, with the western side in general being more non-thermal, with $\alpha \sim -0.7$. As the eastern side is not detected at 1.4GHz (with an rms noise level of ~0.05mJy/beam), that region is likely thermal. [See also Fig. 4 in Marti et al., (2007)]. The flux density in the image ranges from -0.36 mJy/beam to 1.81 mJy/beam. The synthesized beam (resolution) is 13.6" x 13.0" at a position angle of 40 degrees. The figure shown has not been corrected for the mosaic field response, i.e. the fall-off in sensitivity at the edges of the mosaic where the fields do not overlap. Flux measurements were made in an image that was corrected for the mosaic field response (not shown). Contours are plotted from at -3, 2, 4, 6, and 8 × RMS. The synthesized beam is shown in the bottom left corner. A spatial scale bar of 4' or ~ 2pc is shown in the bottom right corner.